\documentclass[pre, twocolumn, footinbib,a4paper, showpacs, superscriptaddress]{revtex4}

\usepackage[latin1]{inputenc}
\usepackage{amssymb}
\usepackage{amsmath}
\usepackage{amsbsy}
\usepackage{bm}
\usepackage{graphicx}
\usepackage{verbatim}
\usepackage{color}
\usepackage{hyperref}

\newcommand{\pe}{\! = \!}

\newcommand{\cE}{\mathcal{E}}

\newcommand{\cw}{c_w}

\def\XXint#1#2#3{{\setbox0=\hbox{$#1{#2#3}{\int}$}
     \vcenter{\hbox{$#2#3$}}\kern-.5\wd0}}

\begin{document}

\title{ Shape and  coarsening dynamics of strained  islands }

\author{Guido Schifani}
\affiliation{Universit\'e  C\^{ot}e d'Azur, CNRS,  INLN, France}
  
\author{Thomas Frisch}
\email{thomas.frisch@unice.fr}
  \affiliation{Universit\'e  C\^{ot}e d'Azur, CNRS,  INLN, France}
    
\author{Mederic Argentina}
  \affiliation{Universit\'e  C\^{ot}e d'Azur, CNRS,  INLN, France}
 \author{Jean-No\"el Aqua}
 \affiliation{Universit\'e  Paris VI, CNRS,  INSP, France}

\begin{abstract}
We investigate the formation and the  coarsening dynamics of islands in a strained  epitaxial  semi-conductor film.  These islands are commonly observed in  thin films  undergoing  a morphological instability due to the presence of  the elasto  capillary  effect.   We first  describe  both analytically and numerically  the formation  of  an equilibrium  island using a two dimensional continuous model. We have found that these  equilibrium  island-like solutions have  a maximum height $h_0$  and they sit on top of a flat wetting layer with a thickness $h_w$.  We then consider two islands and we report that they undergo a non-interrupted coarsening that follows a two stage dynamics. The first stage may be depicted by a quasi-static dynamics, where the mass transfers are proportional to the chemical potential difference of the islands. It is associated with a time scale $t_c$ that is function of the distance $d$ between the islands, and leads to the shrinkage of the smallest island. Once its height becomes smaller than a minimal equilibrium height $h_{0}^{*}$, its mass spreads over the entire system. Our results  pave the way for a future analysis of  coarsening of an assembly of islands.
 \end{abstract}

\maketitle

Understanding the dynamics of coarsening and its effect on self-organisation is a central question in non-equilibrium physics  and solid-state physics since its experimental  discovery by Ostwald at the  end of the 19th century \cite{ostwald}  and  the seminal  theoretical papers of Lishitz-Slyosov and Wagner \cite{lifshitz,wagner} in the late 60's, see also \cite{Marq84}. 
Coarsening is a  general  phenomena  in which the natural size of  a pattern increases with time  in a continuous manner over a large range of time scales \cite{Politi2000271,Ratke2002,opl2010,Biagi2012}.  From a more applied point of view, coarsening has a significant impact on  properties of matter such as the size of grains  in polycrystalline solids, the hardening of metallic alloys, foam dynamics, sintering, sand dunes, etc. We focus here, on the  fundamental aspect of coarsening of strained semi-conductor quantum dots, such as the gallium-aluminum-nitride or silicon-germanium islands \cite{StanHoly04,GrayHullFlor05,Baribeau06,Tu07,McKay09,Berbezier09,Brehm09,jbrault09,ZwanDzur13,BariWu06,Vved08,AquaBerb13,aquaprl2013}. These islands are extensively under scrutiny both for their present and promising applications in electronics or optics, as for example single photons emitters, and for their insights into the fundamental processes of epitaxial growth. The properties and potential applications of  quantum dot assembly are indeed crucially dependent on the amount of coarsening, that may critically affects the size homogeneity of such structures \cite{AquaBerb13}. Moreover, the coarsening of such islands seems to be out of the classical description of Ostwald coarsening and requires more investigation.

The formation of self-organized semi-conductor quantum dots results from the Stranski-Krastanov growth mode \cite{StraKras38}. In this scheme, growth initially proceeds as planar layers, that transform above a given critical thickness $h_c$, into islands separated by a wetting layer. These islands enable a partial relaxation of the elastic stress of the strained film, which overcomes capillary and wetting effects. In SiGe systems, this growth mode includes in fact two different kinetic pathways. The seminal work of Lagally \cite{MoSava90} showed that at large misfit--i.e.~for a large enough Ge composition $x$, in a Si$_{1-x}$Ge$_x$ film, the island growth initiates via the nucleation of large enough fluctuations \cite{EaglCeru90}. On the other hand, at low enough misfit (i.e.~low enough $x$), further experiments \cite{SuttLaga00,TromRoss00} revealed that the island growth begins with a nucleationless instability, reminiscent of the Asaro-Tiller-Grinfeld (ATG) instability \cite{AsarTill72, Grin86,Srol89,SpenVoor91,muller2004}. In this case, the film becomes unstable above the critical height $h_c$, and an initial  surface corrugation increases and transforms after some time into an assembly of quantum dots \cite{CullRobb92,JessPenn93,OzkaNix97,BerbGall98,GaoNix99,SpencerTersoff97,FlorChas99,SuttLaga00,TromRoss00}. After its initial growth, the assembly of islands undergoes some coarsening, driven by the more efficient elastic relaxation of the largest islands. The initial roughly isotropic islands (prepyramids) thence ripen and, as they display steep enough slopes, they transform into anisotropic quantum dots of various sizes, especially pyramids and domes.
Even in the paradigmatic SiGe systems, the nature of the islands coarsening is still a matter of debate and uncertainty \cite{AquaBerb13}. For the initial isotropic islands \cite{PangHuan06,LeviGolo07,AquaFris07}, various theories predict a power-law evolution of the surface roughness and island density at constant mass (annealing), however the exponents of these power laws are clearly different from the classical Ostwald exponents \cite{AquaBerb13}. 
In addition, the coarsening might be impacted by the growth dynamics \cite{AquaFrisVerg10}, the anisotropy of the surface energy  \cite{MedeKami99,ShchBimb03, AquaGouy11, KastVoig99,McKaVena08,AquaFris10,AquaGouy13},  alloying and compositional effects.   

In the present article, we investigate analytically and numerically  the basic but still challenging issue of the coarsening of strained islands in isotropic systems that results from the ATG instability. We have found 
that the island shape can be described by a simple analytical expression and we report the existence   of a continuous  family of solution for the island shape as a function of the  system mass.  Moreover we  have found that the dynamics of coarsening  of two islands can be reduced to a simple two step model.
If the surface evolution might be well described initially in the framework of the linear theory of the ATG instability, the dynamics leads after some time to islands that require a non-linear analysis. The complexity of the dynamics  describing the coarsening of such islands lies in the combination of out-of-equilibrium properties and of the long-range elastic effects. Furthermore, the power law behavior mentioned before arises in the late time dynamics where non-linear effects can not be neglected. We show here that this dynamics is intimately connected to the  static equilibrium shapes of the islands and to the gradient of the chemical potential between two islands.

The article is organised as follows. In the first part, we describe the model under scrutiny that is a 1+1-dimensional strained film that evolves via surface diffusion. In the second part, we characterize analytically the stationary  equilibrium solutions  of  our  model. This solution corresponds to a single island sitting on top of a wetting layer, which characteristics (maximum height $h_0$, surface (or mass) $S$, chemical potential $\mu$) are analytically predicted. In particular, we show that the wetting interactions, yields the existence of a minimal island height. In the third part, we numerically integrate the evolution equation of a simple system composed by two islands with slightly different heights, which interaction leads to a single island after complete coarsening. In the last part, we derive an analytical model that describes the two-islands coarsening dynamics. We show that it is characterized by a two step evolution, with two specific time scales. The first step is well described by a quasi-static approach where each island chemical potential (whose gradient rules the mass transfer between them) is determined by the steady states values. It is associated with an exponential evolution of the islands heights, with a characteristic time scale $t_c$ proportional to the chemical potential gradients, i.e.~to the difference of the islands chemical potentials divided by their separating distance $d$. The second coarsening step occurs once the smallest island is smaller than the minimal stable island height, and therefore quickly dissolves on the wetting layer. It is associated with a second characteristic time scale $\tau$ that describes the dynamics of diffusion of a perturbation on a wetting layer, and that depends on the system size. This two-step dynamical evolution compares favorably with the direct numerical simulation of the coarsening dynamics. The two islands coarsening can be simply modelled by a system of differential equations for each island height. Conclusions and perspectives are drawn in the last part, where this study is promoted in perspective with the more general study of the coarsening of an assembly of islands.


\section{Continuum model}

We study a film-substrate  system, made of a thin film lying on a substrate evolving only via surface diffusion. For studying the formation and the dynamics of the island, we use a standard surface diffusion model which dynamics is governed by \cite{SpenVoor91}:
\begin{equation}
 \frac{ \partial h} {\partial t} = \mathcal{D}  \sqrt{1+ h_x^2} \frac{\partial^2 \mu}{\partial s^2} \   , 
 \label{eqgeneral}
\end{equation}
where $\mathcal{D}$ is the  surface diffusion coefficient, $\partial / \partial s$ the surface gradient and $\mu$ the chemical potential, that depends on the elastic and  the surface energy.
The upper film boundary is free and
localized at $z \pe h(x)$, while the film/substrate interface at
$z=0$ is coherent. We solve the Lam\'e mechanic equilibrium equations with linear isotropic relations. For simplification, we assume that the film and substrate share the same elastic constants. When the film is flat $h(x)=cte$ it is subject to an elastic
stress measured in unit of the volumetric elastic energy
 $\cE_0 = E \, \eta^2/(1-\nu)$.  Here   $\eta \pe (a_f - a_s)/a_s$ is the misfit where  $a_f$ (resp. $a_s$) is the film (resp.
substrate) lattice spacing, $E$ is the Young's  modulus, and $\nu$ the Poisson's coefficient. In the general case, when $h(x)$ displays small slopes, the mechanical equilibrium problem can be solved analytically [see e.g. \cite{AquaFris07}] and its solution is given in terms of the Hilbert transform $\mathcal{H}$ of the surface profile.
In addition, wetting interactions between the film and its substrate prove to be crucial in thin films. It might be describe by a height-dependant surface energy $\gamma(h)$
\cite{Muller96,Spen99,golovin03,muller2003,PangHuan06}. In semiconductor systems, one can consider a smooth $\gamma(h)$ with the generic form characterized by a length
$\delta$, amplitude $\cw$ and generic form $ \gamma (h) = \gamma_f
\left[ 1 + \cw f\left( h/\delta \right)\right]$, where
$f(h\rightarrow \infty)=0$. Here $\delta$ is of
the order of the wetting layer (a few Angstroms).
Adding the elastic and the capillary effects, one finds the chemical potential:
\begin{equation}
 \mu(x) = \cE [h] + \gamma(h) \frac{\partial ^2 h}{\partial x^2} + \gamma'(h)/\sqrt{1+h_x^2}\, ,
\label{eq:defmu}
\end{equation}
where $\cE[h]$ is the volumetric elastic energy on the
surface and the third term in Eq. (\ref{eq:defmu}) is due to wetting where $\gamma '(h)=\frac{\partial \gamma}{\partial h}$. By balancing the elastic energy to the surface energy, we deduce the characteristic length $l_0 \pe \gamma_f/[2(1+\nu)\cE_0] $
describing the typical size  of an horizontal   surface undulation and the associated time scale $t_0=l_0^4/(\mathcal{D}\gamma_f)$.
For example, for a $Si_{0.75} Ge_{0.25}$ film on Si, we find $l_0 = 27 \, {\rm  nm}$ and $t_0 = 23 \,  {\rm s}$ at $700  \, {\rm C}$   (see \cite{chason1990surface} for an estimate of surface diffusion coefficients).
In the small slope approximation, we obtain the following dimensionless equation for the surface evolution
\begin{equation}
\partial_t h=-\partial_{xx}\left[ \partial_{xx}h +\frac{c_w }{\delta}e^{-h/\delta}+\mathcal{H}[\partial_x h] \right] \, ,
\label{dyneq}
\end{equation}
where $\mathcal{H}[\partial_x h]$ is the Hilbert transform  of the spatial derivative of $h(x,t)$  defined as  $\mathcal{F}^{-1}( |k| \mathcal{F}(h))$ where $\mathcal{F}$ is the Fourier transform \cite{AquaFris07}.  
The first term in the r.h.s. Eq. (\ref{dyneq}) represents the stabilising  effect of the surface energy, the second term is the wetting potential and the third term represents the destabilising  effect of the elastic  strain. 
Note that  Eq.  (\ref{dyneq})  represents a conservation equation, and the integral $\int h(x) dx$ (which represents the total amount of deposited material) is constant.
This equation is non-linear, and we use a pseudo-spectral method to solve it numerically \cite{AquaFris07}. Moreover, as we shall see, an analytical insight can be obtained  from an analysis of the stationary solution of Eq. (\ref{dyneq}). 
As shown previously  \cite{AquaFris07}, there exists a critical height $h_c$ above which a flat film becomes unstable  with respect to infinitesimal perturbations,
\begin{equation}
h_c= -\delta \ln(\delta^2/4 c_w).
\end{equation}
For an initial height  above  $h_c$, the initial perturbation evolves towards an assembly of islands that display a non-interrupted coarsening  \cite{AquaFris07} leading to one stationary  island. We describe analytically the characteristics of such a stationary island in next Section.
\section{the stationary island}
\begin{figure}[!ht]
\begin{center}
\includegraphics[width=0.45\textwidth]{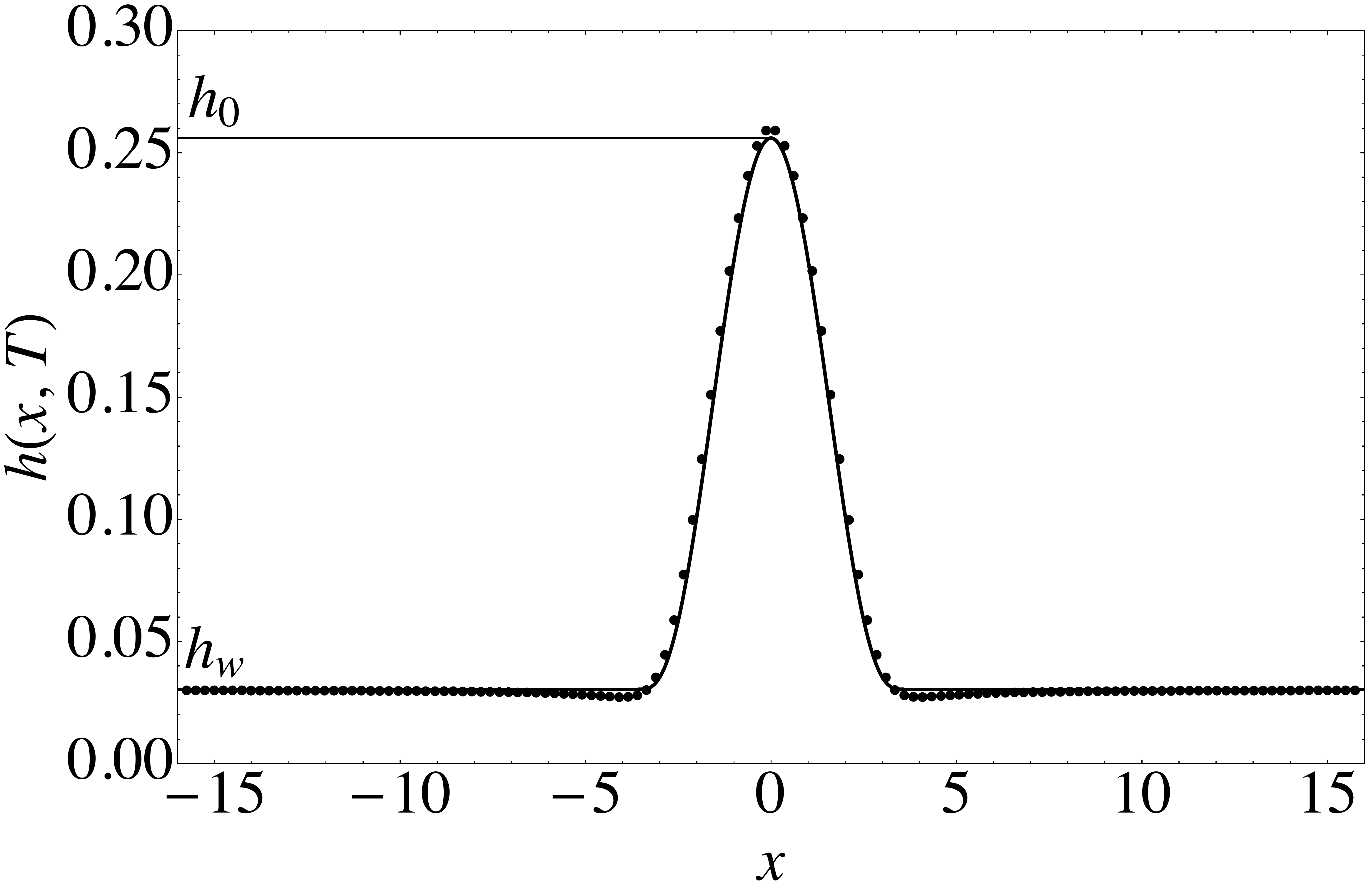}
\caption{\label{fig1} Island like solution resulting from the long time evolution of an initially small surface perturbation. The dots are the stationary profile obtained with numerical simulation of Eq. (\ref{dyneq}).
The system size is $L=32$, $c_w=0.045$ and $\delta=0.005$. The time is $T=1000$. The horizontal and vertical axis are in units of $l_0$.
The line is  the {\it ansatz} given  in Eq. (\ref{ansatz}), with  a width $W= 9 \pi/4$. The value of $h_0$ is taken from  top of the island and  the corresponding value of $h_{w}$ is  obtained from  Eq. (\ref{eqequi}). 
The value of the area  $S = \int_{-L/2}^{L/2} h(x,t) dx=1.5$ is conserved  throughout the dynamics.}
\end{center}
\end{figure}

The goal  of this section is to study the equilibrium stationary solutions of Eq. (\ref{dyneq}), and in particular the island profile. Indeed, above the critical height $h_c$, the evolution of the surface is characterized by a non-interrupted coarsening that eventually leads to a one-island solution \cite{AquaFris07}. This stationary profile is given by one island of height $h_0$ lying on top of a wetting layer of thickness $h_w$, see Fig \ref{fig1}. It is characterized by a constant chemical potential $\mu$ on the surface,
\begin{equation}
\mu=-\partial_{xx}h -\frac{c_w }{\delta}e^{-h/\delta}-\mathcal{H}[ \partial_x h] \, .
\label{equationequilibre}
\end{equation}

The  stationary  island characteristics maximum height $h_0$ and width $W$, can be predicted by the use of a simple model. This model has a no free parameters  and can be characterised by the total surface of the system $S= \int_{-L/2}^{L/2} h(x,t) dx$ with $L$ the system size. Thus islands of different height $h_0$ can be generated numerically by varying the control parameter $S$ in the initial condition.
 \begin{figure}[!ht]
\begin{center}
\includegraphics[width=0.45\textwidth]{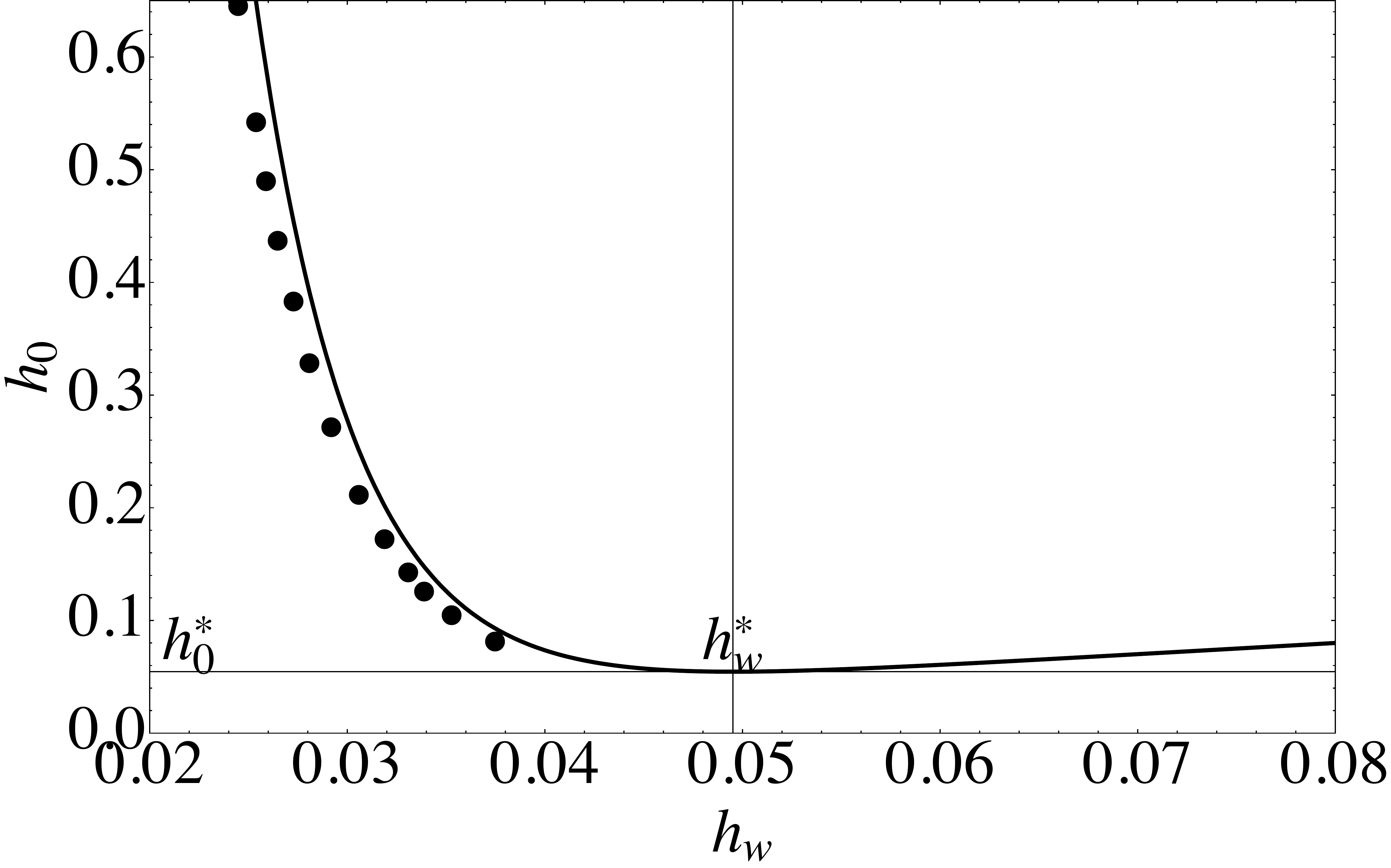}
\caption{\label{fig2}  Height of the island $h_0$ as a function of $h_{w}$ in units of $l_0$. Dots are obtained by simulations of Eq. (\ref{dyneq}) and the full line is the {\it ansatz} given in Eq. (\ref{ansatz}). The value of $h_{0}^{*}$ is defined on the figure. The different point are obtained by  performing different simulations for different value of the initial surface $S$. The value of the parameters $L$, $c_w$ and $\delta$ are the same then the one used in Fig. 1. The mimal value of $h_{0}^{*}$ will be defined in Eq. (\ref{hmin})}.
\end{center}
 \end{figure}
 Motivated by the result of the numerical simulation of Eq. (\ref{dyneq}),  we choose the following {\it ansatz} for the stationary solution of Eq. (\ref{dyneq}). For $|x|<W/2$
\begin{equation}
h(x)=(h_0-h_{w}) \left( \frac{2}{W}\right)^6\left[\left(\frac{W}{2}\right)^2-x^2\right]^3+h_{w}  \, ,
 \label{ansatz}
\end{equation}
while  for $|x|>W/2$, we choose   $h(x)=h_{w}$. This {\it ansatz} satisfies  the continuity of the function at $|x|=W/2$  and of its derivatives up to third order and is characterized by three parameters. After substitution of this {\it ansatz} in Eq.  (\ref{equationequilibre}),  and  using a simple polynomial expansion around the point $x=0$ up to  second order in $x$, we obtain at order $x^0$ the following relation between the island height $h_0$ and the height of the wetting layer $h_w$:  

\begin{equation}
h_0=h_{w}+\frac{135 \pi^2 }{128}\frac{c_w}{\delta}e^{-h_{w}/\delta} \, .
 \label{eqequi}
\end{equation}
At order $x^2$, we obtain the relation for the width of the island
$W=\frac{9\pi}{4}$ \footnote{The calculation of the Hilbert transform  is  done in  real space using the standard definition of the principal value integral.}.

In  Fig. \ref{fig1},  we compare the profile of a stationary island obtained by numerical simulation of Eq.  ($ \ref{dyneq}$) with this {\it ansatz}. The agreement between the two is rather good with small discrepancies located on a small zone
 at the foot of the island \footnote{These discrepancy can be   improved by using higher order polynomial or  matching methods between the wetting layer and the islands. 
 However, an improvement of the solution does not lead to any qualitative change}.

We also plot in Fig. \ref{fig2} the height of the island $h_0$ at equilibrium as a function of the height of the wetting layer far away of the island $h_{w}$.  The simulation values are obtained by varying the system surfaces ($S$) while the {\it ansatz} result follows from Eq. (\ref{eqequi}). Again, the agreement is rather good. Of special interest is the fact that $h_0$ has a minimal value called $h_{0}^{*}$
The critical height  $h_{0}^{*}$  is defined by the relation $ \frac{\partial h_0}{\partial_{h_{w}}}=0$, this leads using Eq. (\ref{eqequi}) to the result: 
\begin{equation}
h_{0}^{*}=\delta\left[1+ \ln{ \left(\frac{ c_w 135 \pi^2}{\delta^2 128}\right)}\right] \, ,
\label{hmin}
\end{equation}
while the associated wetting thickness is:
\begin{equation}
h_{w}^{*}=\delta\ln{ \left(\frac{ c_w 135 \pi^2}{\delta^2 128}\right)} \, .
\label{hwm}
\end{equation}
As we observed numerically, islands with  $h_0$  smaller then $h_{0}^{*}$  are not stable. 
Hence, the presence of wetting interactions enforce the existence of minimal value of the equilibrium island surface, in addition to the existence of a minimal film thickness $h_c$. 
The critical island height could be observed experimentally  
and it will be important in the description of the coarsening process. 

As regards to the chemical potential, each island-like stationary solutions of Eq.  (\ref{equationequilibre}) is defined by
\begin{equation}
\mu_{i} =-\frac{c_w}{\delta} e^{ -h_{w}/\delta}\, .
\label{mu}
\end{equation} 
This results comes from the fact   that far from the island the film is rather flat so that $h_x$ and $h_{xx}$ vanish and only the wetting potential term remains  dominant in Eq. (\ref{equationequilibre}).
Therefore the simple knowledge of $h_w$ can leads to the determination of the chemical potential and reciprocally.
Using Eqs. (\ref{hwm}) and  (\ref{mu}), we find that the critical chemical potential $\mu^{*}$ associated with the critical solution with $h_{0}^{*}$ reads
 \begin{equation}
 \mu^{*}= -\delta\frac{128}{135  \pi^2} \, .
 \label{muii}
\end{equation}

We mentioned  previously  that islands are uniquely characterised by the surface $S$. Now that we have the profile of the island given in Eq. (\ref{ansatz}), we can calculate its  the surface $S$, 
  \begin{equation}
 S=h_{w} L+\frac{243 \pi ^3   }{224  }\frac{c_w}{\delta} e^{-h_{w}/\delta }  \equiv \langle h \rangle L \ .
 \label{surface1}
\end{equation}
The total surface (mass) $S$ can thus be varied by varying the mean height   $\langle h \rangle$ or the size $L$ of the system.
\begin{figure}[!ht]
\begin{center}
\includegraphics[width=0.45\textwidth]{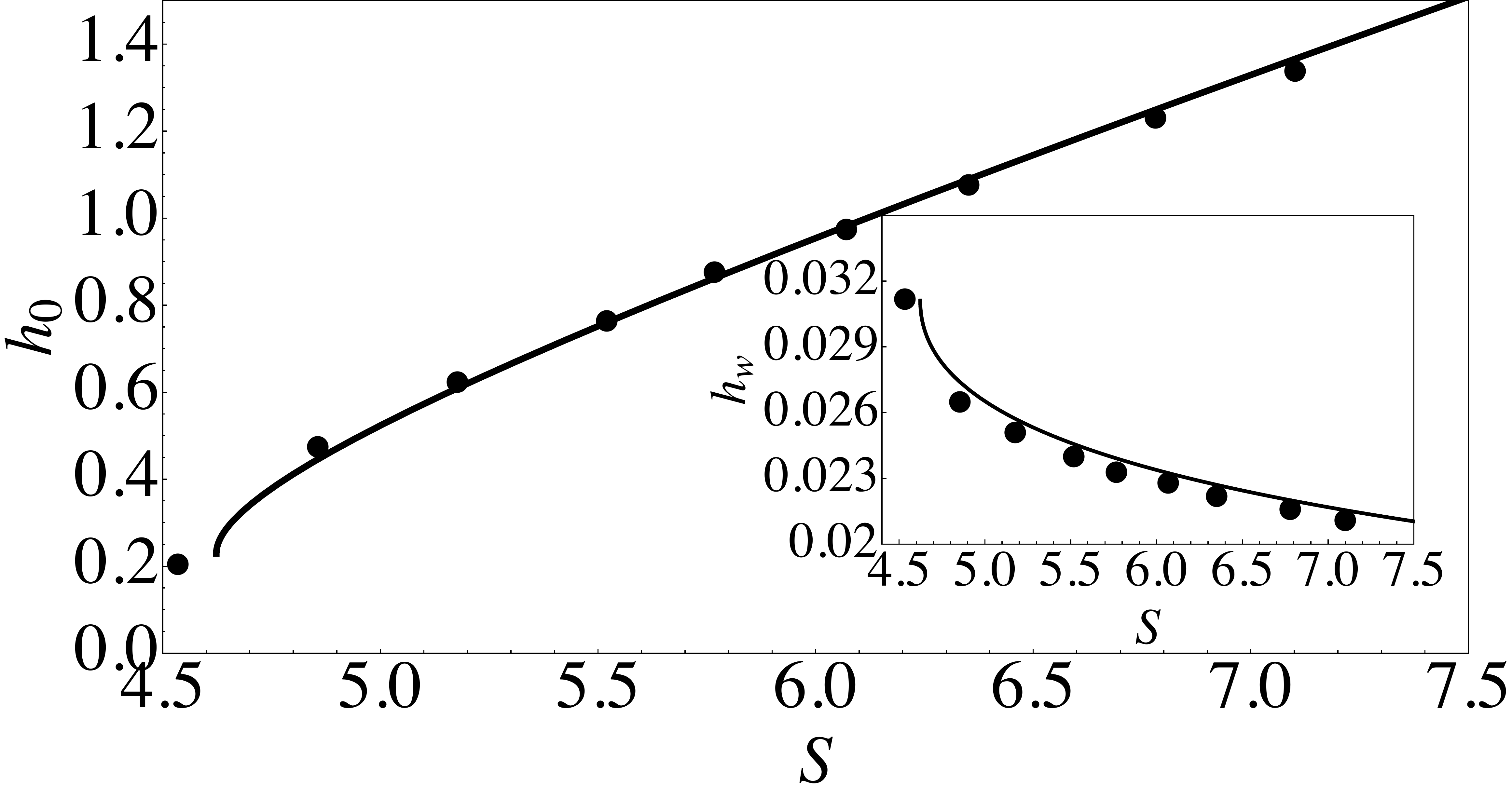}
\caption{\label{fig3} The  height $h_0$ as a function of the surface $S= \langle h \rangle L$ with $L$ being fixed. The horizontal and vertical axis are in units of $l_0^2$ and $l_0$ respectively. The dots are obtained by numerical simulation of  Eq. (\ref{dyneq}).
 The curve  corresponds to  Eq. (\ref{surface1}) and Eq. (\ref{eqequi}). The inset is the height $h_w$ as a function of $S$. The system size is $L=128$, $c_w=0.045$ and $\delta=0.005$.}
\end{center}
\end{figure}

We plot in Fig. \ref{fig3}, the island maximum height $h_0$  and the height of the wetting layer $h_{w}$ versus the surface $S$ by varying $\langle h \rangle$. As expected, we observe in Fig. \ref{fig3} that the maximum height of the island increases as the surface $S$ increase. 
As $h_0$ is a decreasing function of $h_w$, see Fig. \ref{fig2}, we also find that $h_w$ is decreasing function of the island surface $S$ as shown in the inset of Fig. \ref{fig3}. This may be associated with the larger relaxation of the larger islands that are in equilibrium with a more stable thin wetting layer.
\begin{figure}[!ht]
\begin{center}
\includegraphics[width=0.45\textwidth]{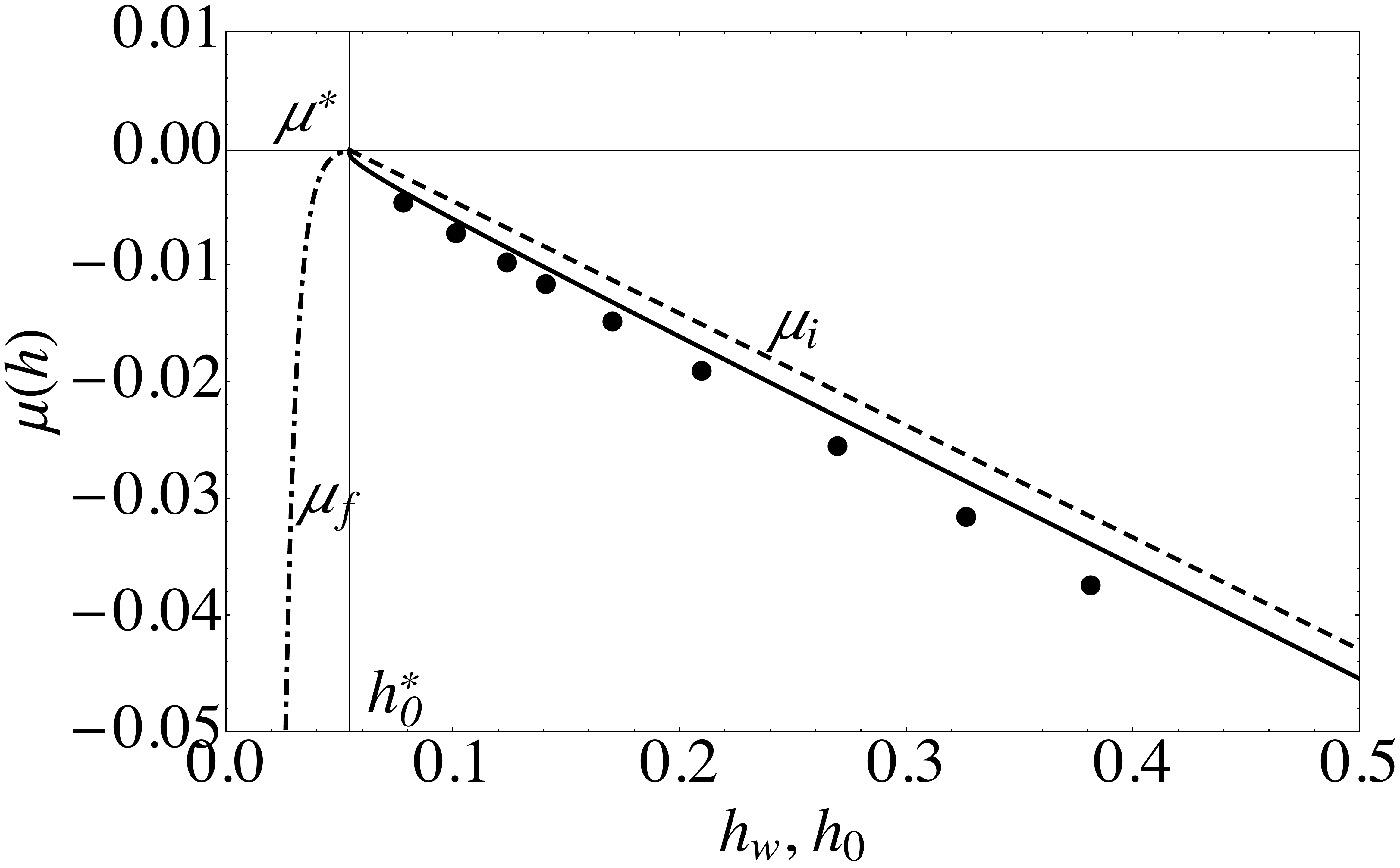}
\caption{\label{fig4}  For  $h<h_{0}^{*}$, the dash-doted line is the chemical potential $\mu=-\frac{c_w}{\delta} e^{-h/\delta}$  as a function height for the flat film. The unit of the vertical axis is in $\cE_0 = E \, \eta^2/(1-\nu)= 6.7*10^7  {\rm Joules}/{\rm m}^3$ and the unit of the horizontal axis is in $l_0$.
For $h>h_{0}^{*}$,  the horizontal axis $h=h_0$. The dots represent the numerical simulation for the equilibrium state of an island given by Eq. (\ref{dyneq}). The continuous curve is the prediction  given using  Eq. (\ref{eqequi}) and Eq. (\ref{mu}) for the chemical potential of the island. The dashed curve is given by Eq. (\ref{mulin}).}
\end{center}
 \end{figure}
 
We now study the chemical potential associated with the one island solution. For $h_0\geqslant h_0^{*}$, there exists an equilibrium island solution. Its chemical potential is given by Eq. (\ref{mu}) in terms of the wetting layer thickness $h_w$. The equilibrium island chemical potential is plotted as a function of $h_0$ in Fig. \ref{fig4}. As the island surface increases, hence $h_0$ increases, the island chemical potential naturally decreases, showing the larger elastic relaxation of large islands. This conclusion was also found in the three-dimensional island under study in \citep{AquaFris07}. When $h_0<h_0^{*}$, only the flat film solution exists, its chemical potential is entirely given by Eq. (\ref{muii}). We also plot this chemical potential as a function of $h_w$ in Fig. \ref{fig4}. It is an increasing function of $h_w$ as enforced by the (attractive) wetting interactions. At equilibrium, for $h>h_c$, an island of thickness $h_0$ coexist with a wetting layer of thickness $h_w$, that have the same chemical potential. In Fig \ref{fig4}, we again find a good agreement between the numerical simulation and our theoretical prediction. As expected the chemical potential has a maximum value  $\mu^{*}$, given by Eq. (\ref{muii}),  associated with the minimal value of the surface height $h_{0}^{*}$. The dashed  curve in Fig. \ref{fig4} represents  the linear approximation to  $\mu_i$,
  \begin{equation}
 \mu_{i}^{l} \simeq -c (h_0-h_{0}^{*}) +\mu^{*}\, ,
 \label{mulin}
 \end{equation}
that has been obtained using  Eq. (\ref{eqequi}) and Eq. (\ref{mu}), here $c=\frac{128}{135 \pi^2}$. 

\section{ Coarsening of two islands}
\begin{figure}[!htb]
\begin{center}
\includegraphics[width=0.45\textwidth]{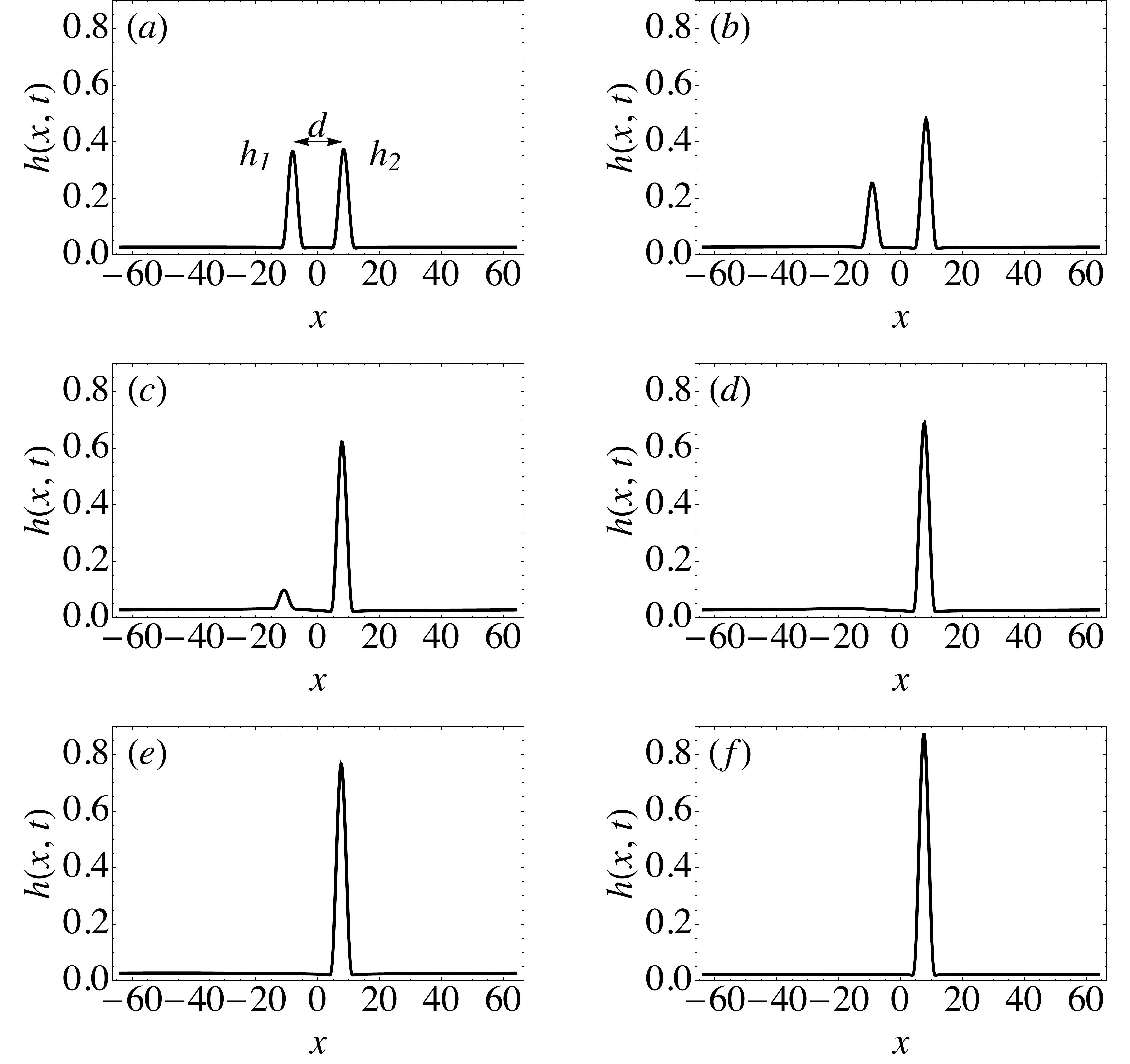}
\caption{\label{fig5}    Numerical  resolution of Eq. (\ref{dyneq}) for the profile evolution of two interacting islands separated by a distance $d$. The horizontal and vertical axis are in units of $l_0$. The system size is $L=128$. The initial condition consists of two islands separated by a distance $d=16$ and initial amplitudes $h_1=0.36$ (left island) and $h_2=0.37$ (right island )with time
a) $t=0$, 
b)  $t=700$ , 
c) $t=1080$ before $t_c$,
d) characteristic time $t=t_c=1350$, 
e) $t=1550$ and 
f) $t=2580$ when the equilibrium state is reached.}
\end{center}
\end{figure}

We now address the question of  coarsening of two islands of slightly different  amplitudes (heights) separated by a distance $d$.
Let $h_1$ and $h_2$ be the height for the small and large islands respectively (left peak and right peak on Fig. \ref{fig5}). These quantities will evolve with time. In Fig. \ref{fig5}, we represent the time  evolution of the two islands as enforced by the dynamical evolution Eq. (\ref{dyneq}).  
The initial condition is given by two islands at equilibrium with amplitude $h_{1}=h_i - \epsilon$ and $h_{2}=h_i + \epsilon$. We find a first regime where the height of the small island decreases while the  height of the  large  island increases. Then, the small island reaches the critical height  $h_{0}^{*}$ at time $t_c$ (Fig. \ref{fig5}.d).  In the second regime for $t>t_c$ (Fig. \ref{fig5}.e), the remaining mass in the wetting layer   diffuses  towards the larger island, which relaxes towards its equilibrium state (Fig. \ref{fig5}.f). The largest island height $h_2$ constantly increases during the whole coarsening process.  

\begin{figure}[!htb]
\begin{center}
\includegraphics[width=0.45\textwidth]{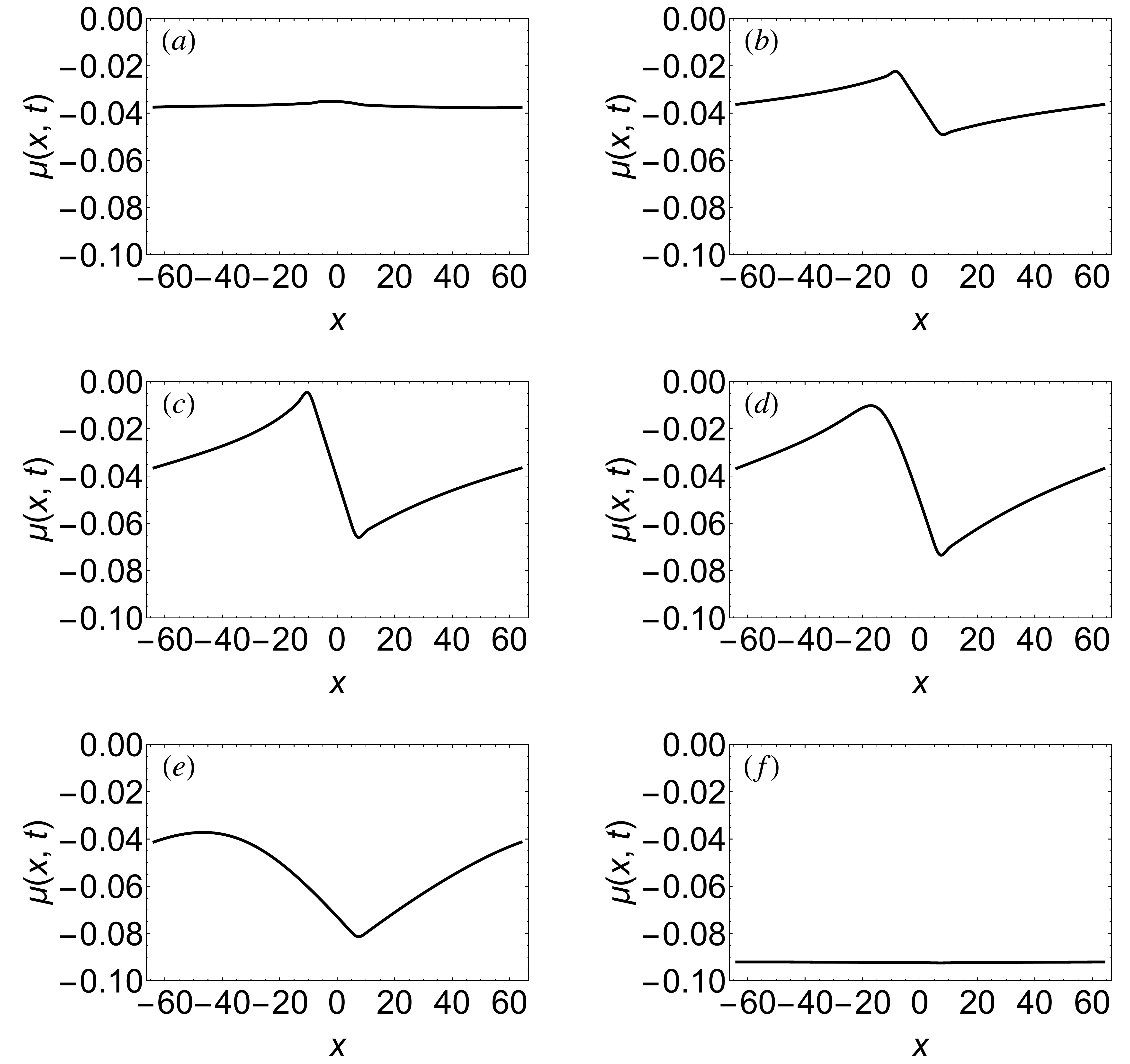}
\caption{Numerical evolution of Eq. (\ref{dyneq}) for the chemical potential of two interacting islands corresponding to Fig. \ref{fig5}. The unit of the vertical axis is $\cE_0= 6.7*10^7  {\rm Joules}/{\rm m}^3$. The horizontal axis is in unit of $l_0$.}
\label{fig6}
\end{center}
\end{figure}

In Fig. \ref{fig6}, we plot the temporal evolution of the local chemical potential associated with the evolution given by Eq. (\ref{dyneq}). 
The chemical potential on the small island increases when  its  height decreases as it becomes less and less stable, and  conversely for the   large island.  Before $t_c$, the chemical potential $\mu$ between the two islands is a linear decreasing function of space as shown for example in Fig. \ref{fig6}.b  and Fig. \ref{fig6}.c. Furthermore, when $t<t_c$, and outside the islands,  the chemical potential has variations on the scale of the system $L$. It is due to finite size effects that can be neglected as long as $d<<L$. When the critical height of the small island is reached (Fig. \ref{fig6}.d) at time $t=t_c$, the chemical potential  of the small island is equal to $\mu^{*}$ and the height of the small island $h_1$ is $h_{0}^{*}$. For $t>t_c$, while $h_2$ is growing, the diffusion on the wetting layer takes place on a scale of the order $L$.
This second regime relaxes towards equilibrium, where finally the chemical potential is constant, Fig. (\ref{fig6}.f).

\section{model of coarsening}

We now develop a simple mean-field model that describes the coarsening phenomena  in two stages. In this model the islands are represented by a punctual object of varying surface. The advantage of this model is that it requires only a small number of input parameters such as the width of the island $W$ and the chemical potential difference between the two islands. We make the assumption that the dynamics is close to equilibrium so that the results for stationary island can be exploited.
 The first coarsening stage is defined for $t<t_c$  when the two islands co-exist while for $t>t_c$,  the smaller island has disappeared and perturbation of the wetting layer diffuses  towards the larger island.

\begin{figure}[!ht]
\begin{center}
\includegraphics[width=0.45\textwidth]{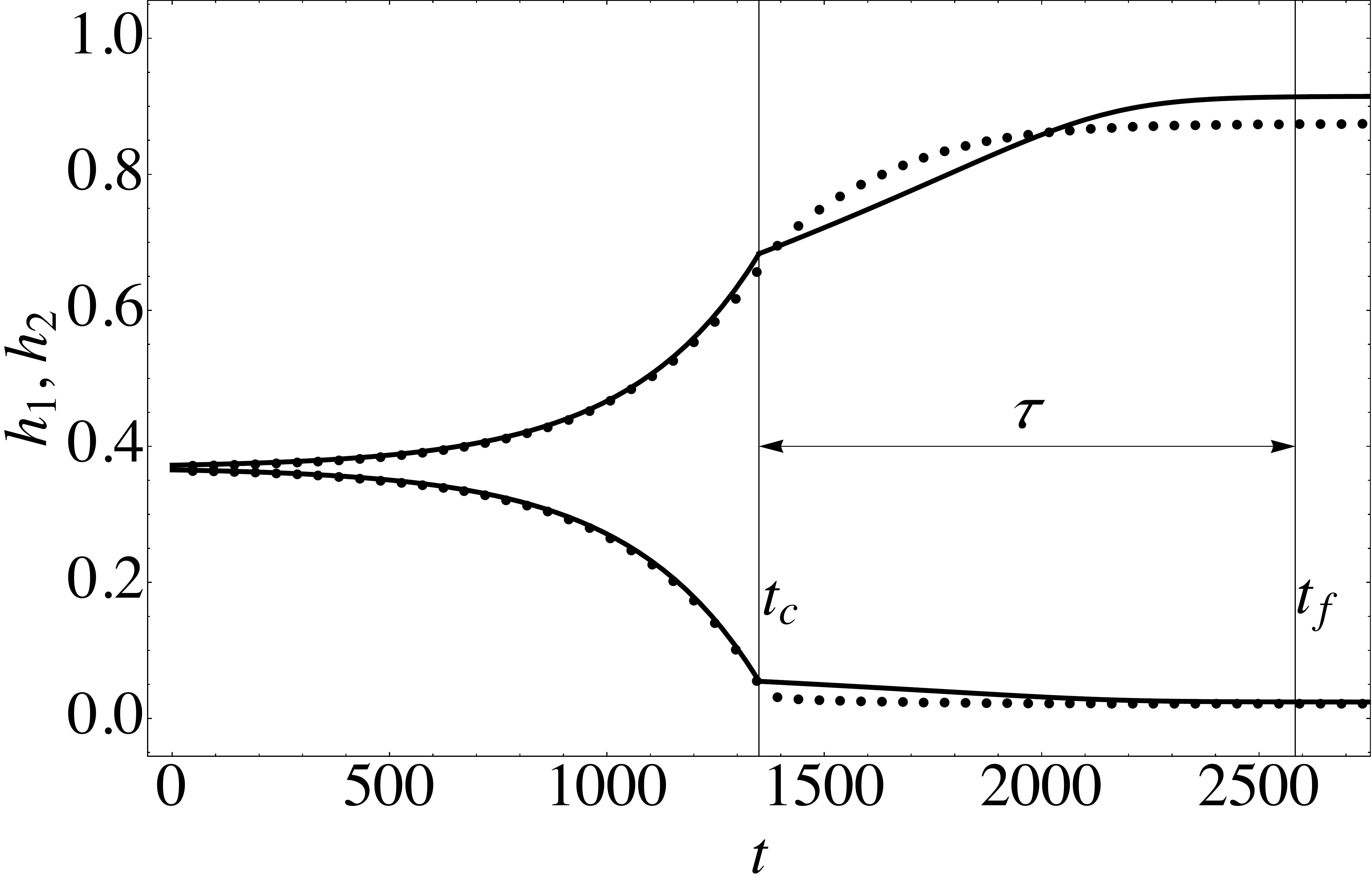}
\caption{\label{fig7}      Amplitude $h_1$ and $h_2$ of the islands as a function of time. Full curves are the theoretical prediction and the dashed curve is the numerical simulation. The times $t_c$ and $t_f$ are represented on the figure.  $\tau$ is defined as the time since $t_c$  for which  the  amplitude $h_2$ of the large island  has reach 0.99 of its equilibrium value. The horizontal and vertical axis are in units of $t_0$ and $l_0$ respectively.}
\end{center}
\end{figure}

For $t<t_c$, we model  the  dynamics of the height of each island  based on the flux of matter induced by the chemical potential gradient between the two islands.  
This  spatial gradient takes place on a  length scale of order $d$. Mass conservation enforces in this approximation \footnote{For simplicity, we neglect finite size effects which lead to  small terms in $d/L$ due to the  presence of periodic boundary conditions.}
\begin{equation}
\begin{array}{ll}
\alpha W \partial_t h_1=\frac{\mu_{i}(h_2)-\mu_{i}(h_1)}{d}\\
\alpha W \partial_t h_2=\frac{\mu_{i}(h_1)-\mu_{i}(h_2)}{d}
\end{array}
,
\label{eq:dyn-mod}
\end{equation}
where $h_1$ is the height of the small island, $h_2$ the height of the large one, $W$ their width and $\alpha$  a constant  geometrical factor which is of order  one \footnote{$\alpha=\int_{-W/2}^{W/2} h(x) dx/ h_0 W=0.4636$ and $\beta=0.22$.}. 

Furthermore, we assume in the following that the island chemical potential might be given by the linear form given in Eq. (\ref{mulin})
Hence, the system (\ref{eq:dyn-mod}) simplifies into
\begin{equation}
\begin{array}{ll}
 \alpha W \partial_t h_1=-\frac{ c(h_2-h_1)}{d}\\
 \alpha W \partial_t h_2=-\frac{ c(h_1-h_2)}{d}
\end{array}
,
\label{firststage}
\end{equation}
where $c=\frac{128}{135 \pi^2}$, given by the slope of Eq. (\ref{mulin}). Let us write the amplitude of the islands 
\begin{equation}
\begin{array}{ll}
h_1(t)=h_i-\epsilon \tilde{h}(t)\\
h_2(t)=h_i+\epsilon \tilde{h}(t)
\end{array}
,
\label{eq:dyn-mod-2}
\end{equation}
which implies that $h_1(t)+h_2(t)= 2 h_i$ and $\tilde{h}$ is the perturbation of the stationary state. Solving (\ref{firststage}), we deduce  that the perturbation increase exponentially
\begin{equation}
\tilde{h}(t)=e^{\frac{ 2c}{d \alpha W}t},
\end{equation}
in the first temporal regime. This regime extends up to $t_c$, such as  $h_1(t_c)=h_{0}^{*}$ which  leads to $h_{0}^{*}=h_i-\epsilon e^{ \frac{ 2c}{d \alpha W}t_c}$.  Hence, we find
\begin{equation}
t_c=\frac{d\alpha W}{2 c}\ln\left[ \frac{h_i-h_{0}^{*}}{\epsilon}\right].
\label{eq:tc}
\end{equation}
As shown on Fig. \ref{fig7}, there is a good agreement between the numerical simulation and this estimate.

\begin{figure}[!ht]
\begin{center}
\includegraphics[width=0.45\textwidth]{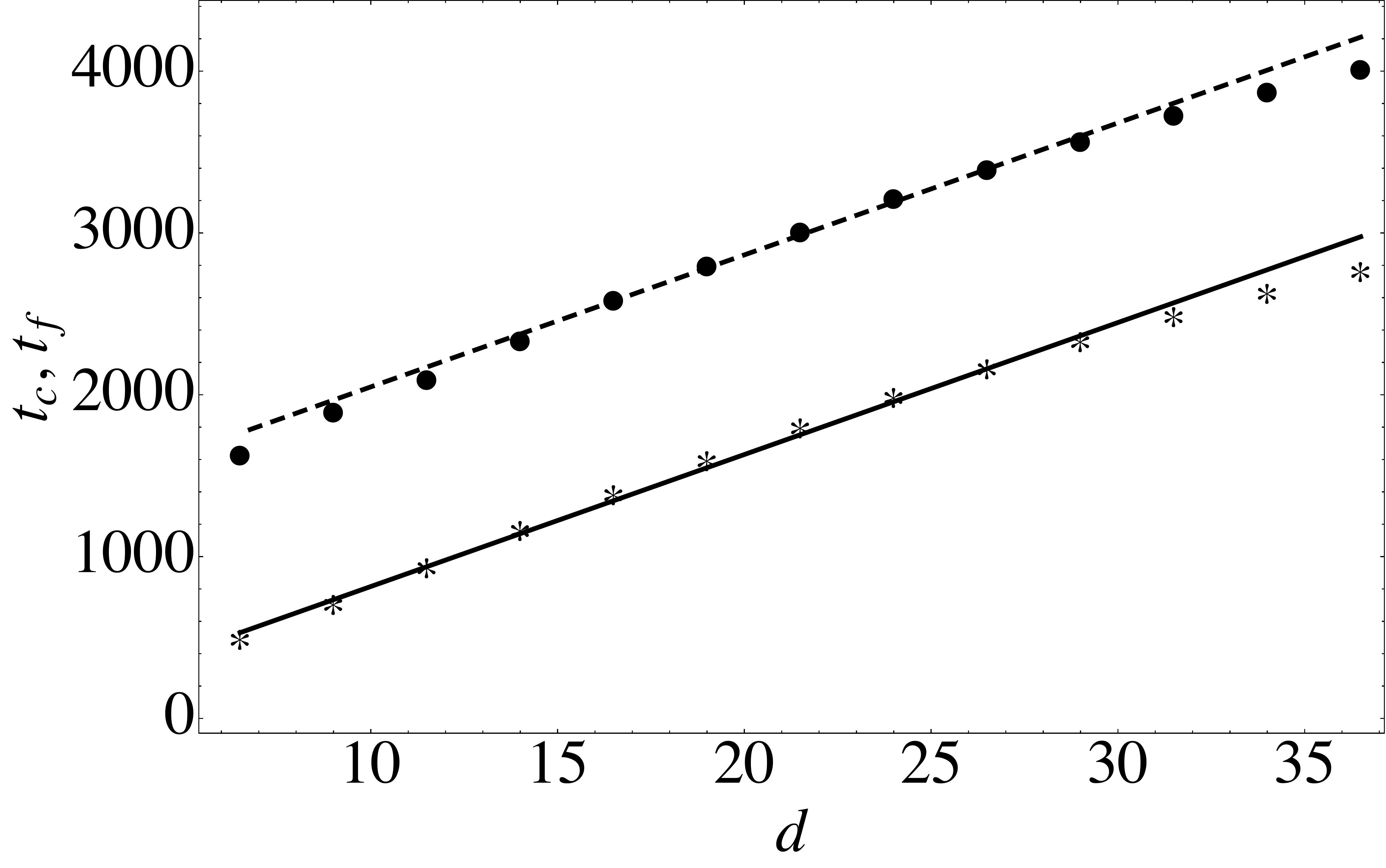}
\caption{Characteristic times $t_c$ and $t_f$  ($*$ and $\bullet$ respectively), as a function of the distance $d$ between the islands, obtained by numerical simulation. The line is $t_c$ from Eq. (\ref{eq:tc}) and the dashed line is the $t_f+\tau$, where $\tau$ is obtained with the numerical solution of Eq. (\ref{eqh2}). The system size is $L=128$. The time $t_f$ for the disappearance of the two islands increases with the system size,  it is linear when $d/L\ll1$. When d increases and becomes of the order of L there are deviation from the linear law due to the effect of the periodic boundary conditions.  The horizontal and vertical axis are in units of $l_0$ and $t_0$ respectively.}
\label{fig8} 
\end{center}
\end{figure}

\begin{figure}[!ht]
\begin{center}
\includegraphics[width=0.45\textwidth]{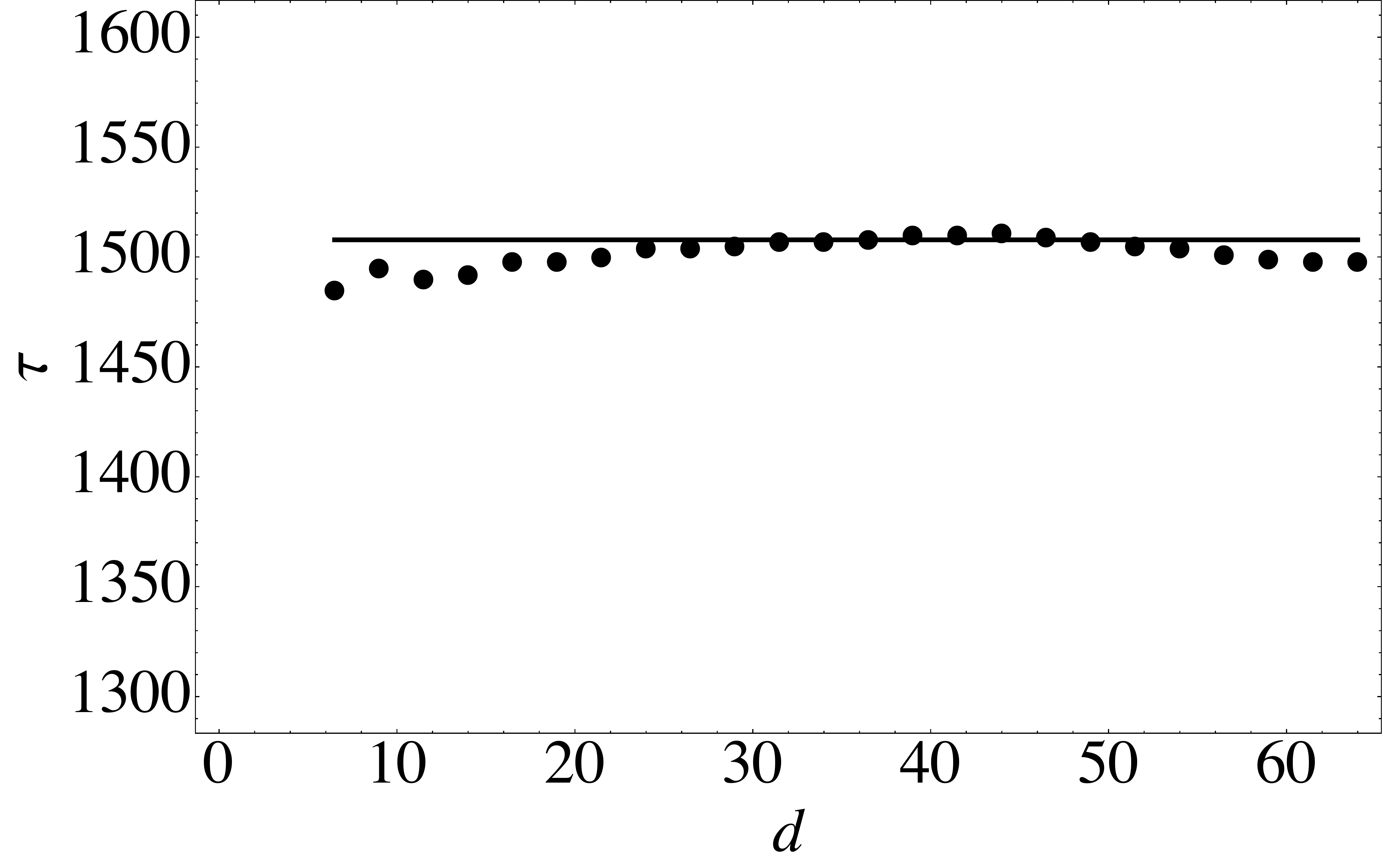}
\caption{ Characteristic time $\tau$  as a function of the distance $d$ between the islands, obtained by numerical simulation of Eq. (\ref{dyneq}). The line is the time $\tau$ obtained with the solution of Eq. (\ref{eqh2}). The horizontal and vertical axis are in units of $l_0$ and $t_0$ respectively.}
\label{fig9}
\end{center}
\end{figure}
The second regime is reached when the amplitude of the small island becomes smaller than the critical height $h_{0}^{*}$, $h_1<h_{0}^{*}$ at $t>t_c$.
Mass diffusion then occurs on the wetting layer. The characteristic time   $\tau$  of this second regime then depends essentially on the full size of the system $L$ and only weakly on the distance $d$.
To quantify, we  write the mass conservation equation as 
\begin{equation}
\beta (L- W) h_1 +\alpha W h_{2}= S ,
\end{equation}
where  $\alpha$ and $\beta$ are geometrical factors  respectively for the island and for the wetting layer while $S$ is fixed by the initial conditions. From this relation, we deduce that 
\begin{equation}
\partial_t  h_1=-A \partial_t  h_2   \,  \quad  \text{here} \quad \,  A=\frac{\alpha W}{\beta (L-W)}.
\label{eqh1}
\end{equation}
Again, we have assumed that the growth rate of the island is proportional to the gradient of chemical potential. This gradient occurs on a scale of order $L$ so that
\begin{eqnarray}
\alpha W \partial_t h_2 = \frac{2 [\mu_f(h_1)-\mu_{i}^{l}(h_2)]}{L}.
\label{eqh2} 
\end{eqnarray}
Here $\mu_f(h_1)=-\frac{c_w}{\delta} e^{-h_1/\delta}$ is the approximate wetting chemical potential of the wetting layer.
In order to obtain the time evolution of $h_1(t)$ and $h_2(t)$, we have integrated numerically Eqs. (\ref{eqh1},\ref{eqh2}). 
As shown on Fig. \ref{fig7}, the system  of  Eqs. (\ref{eqh1},\ref{eqh2}) captures well the numerical evolution of Eq. (\ref{dyneq}). The amplitude of  the island increases  with time before saturating at a value  close to the predicted value which depends on the value of $S$ as shown in  Fig \ref{fig3}.

In order to  quantify  this coarsening  process, we define the  time $t_f$ as the time at which  the amplitude of the large island  has reached 0.99 of its equilibrium value. In addition, we define $\tau$ such as $t_f  =(\tau+t_c)$.


In Fig. \ref{fig8}, we plot  the different times $t_c$ and $t_f$ as a function of the distance $d$ between  the islands using the numerical and the analytic results Eq. (\ref{eq:tc}). We observe, as long as $d/L$ is small,  that $t_c$ increases linearly  with the distance $d$ as predicted by Eq. (\ref{eq:tc}).   When $d$ increases and becomes of the order of $L$ there are   deviation from the linear law in $d$ due to the image interaction  since our numerical simulation are performed in 
a periodic system.  In  Fig. \ref{fig9}, we show that the time  $\tau  $  is almost independent of the distance $d$ separating the islands.
As a conclusion, Figs. \ref{fig8} and \ref{fig9} show that $\tau$ is independent of $d$ while $t_f$ and  $t_c$ increase linearly with $d$.

\section{Conclusion and perspectives}
We have studied in this article the  dynamics and the  coarsening of strained islands.  We first obtained an approximate analytical equation for a stationary island lying on a wetting layer. This approach allows to predict the width  $W$ of the island and  to  relate the island amplitude  to the height of the wetting layer. We  have shown that  the  presence of the wetting  potential  leads to the existence of a critical island height $h_{0}^{*}$ below which the island does not exist.  
The comparison between the approximate analytical solution and the stationary state resulting from the numerical integration of the mass diffusion equation is good.
Secondly, we have  investigate the dynamics of coarsening of  two islands and we have found that  this coarsening is non-interrupted, the small island disappears in favour of the largest one.
As observed numerically,  in a first regime the height of the largest island  increases exponentially until a time $t_c$ at which the smallest island becomes unstable. The characteristic time $t_c$ scales like the distance $d$ between the islands. In a second regime, which lasts a time $\tau$,  the perturbation on the wetting layer diffuses and the amplitude of the  remaining island grows until its reaches its equilibrium value.
This second regime is quite independent of the distance $d$  between the initial island.
In order to model this dynamics, we propose a simple model based on a quasi-static hypothesis with mass currents driven by the gradient of the chemical potential. 
These results pave the way for a description of coarsening in strained systems with long-range interactions.
 We will extend this analysis to the problem of coarsening  of an array of $N$  islands as generated by the Asaro-Tiller-Grinfeld instability by generalizing the set of  Eqs. (\ref{eq:dyn-mod}) to  N  islands.
 An extension of this  analytical work on three dimensional islands  with inclusion of the surface energy anisotropy will be considered in the future. 
\begin{acknowledgements}
We would like to thank  Pierre M\"uller, Julien Brault, Benjamin Damilano, Philippe Venn\'egu\`es, Matthieu Leroux, Jean Massies, Franck Celestini, Jean Rachenbach,   Isabelle Berbezier and  Alberto Verga  for useful discussion. We thank the ANR NanoGanUV for financial support.
 \end{acknowledgements}

\end{document}